\documentclass[11pt]{article}

\usepackage[letterpaper,margin=1in]{geometry}
\usepackage{amsmath,amssymb,mathtools}
\usepackage{graphicx}
\usepackage{bm}
\usepackage{booktabs,longtable,array}
\usepackage{microtype}
\usepackage{xcolor}
\usepackage[colorlinks=true,allcolors=blue!55!black]{hyperref}

\DeclareMathOperator{\haf}{haf}

\newcommand{\E}{\mathbb E}
\newcommand{\C}{\mathbb C}
\newcommand{\R}{\mathcal R}
\newcommand{\N}{\mathcal N}

\newcommand{\LeanDecl}[1]{\nolinkurl{#1}}
\begin{document}

\title{Exact Moments of Gaussian Gram Hafnians Reveal an $n^2/\log n$ Threshold for Weak Anticoncentration}

\author{Hongru Zhao\\[0.25em]
\small School of Statistics, University of Minnesota, Minneapolis, Minnesota 55455, USA\\
\small \texttt{zhao1118@umn.edu}}
\date{}

\maketitle

\begin{abstract}
Anticoncentration is central to approximate sampling hardness arguments. In
the independent Gaussian surrogate for collision free Gaussian boson sampling,
the moment ratio analyzed below also enters the averaged ideal linear cross
entropy reference value. Let $H_{k,n}=\haf(X^{\mathsf T}X)$, where
$X\in\C^{k\times2n}$ has independent
standard circular complex Gaussian entries. We evaluate
$\E|H_{k,n}|^2$ and $\E|H_{k,n}|^4$ exactly by reducing four hafnian copies
to a rank two Gaussian integral. For
$\R_{k,n}=(\E|H_{k,n}|^2)^2/\E|H_{k,n}|^4$, we obtain
$\R_{k,n}=4^{-n}\binom{2n}{n}/F_{k,n}$, where
$F_{k,n}={}_3F_2(-n,-n,1/2;1,k/2;1)$ is a terminating generalized
hypergeometric polynomial. If $k/n^2\to c>0$, then
$F_{k,n}\to e^{1/c}I_0(1/c)$, where $I_0$ is the modified Bessel function of
the first kind of order zero, and consequently
$\R_{k,n}\sqrt{\pi n}\to[e^{1/c}I_0(1/c)]^{-1}$. Thus $k\asymp n^2$ is a
smooth Bessel crossover, whereas the scaling order boundary for inverse
polynomial weak anticoncentration is $k\asymp n^2/\log n$. These conclusions
concern the Gaussian surrogate moment criterion; finite dimensional Haar
moment transfer and high probability small ball anticoncentration remain
separate problems.
\end{abstract}

Hardness of evaluating a photonic output probability is not, by itself,
hardness of sampling the full output distribution. In the standard reduction,
Stockmeyer counting applied to a hypothetical classical sampler estimates the
sampler probability $q_U(S)$ multiplicatively. Total variation closeness and
hiding then give additive control relative to the ideal probability $p_U(S)$;
anticoncentration converts this control into relative information on a
nonnegligible set of instances
\cite{AaronsonArkhipov2013,HangleiterEisert2023}. In collision free Gaussian
boson sampling (GBS), the relevant weights are absolute squares of hafnians of
dependent Gram type matrices rather than absolute squares of permanents of IID
matrices \cite{Hamilton2017,DeshpandeEtAl2022}. Their second moment ratio is
therefore a tractable diagnostic for this reduction and an
ideal linear cross entropy reference quantity. Uniform small ball control with
arbitrarily high success probability is a stronger target.

Consider $k$ squeezed input modes and a collision free event with $2n$
detected photons. In the independent Gaussian surrogate, its matrix factor is
$H_{k,n}=\haf(X^{\mathsf T}X)$ for $X\in\C^{k\times2n}$. Ehrenberg
\emph{et al.} developed a graph expansion for the fourth moment
\cite{EhrenbergPRL2025} and, in their companion work
\cite{EhrenbergPRA2025}, an exact recursion evaluated through $2n=80$. On the
basis of those results, they conjectured that weak anticoncentration fails for
$k=O(n^2)$ and holds for $k=\omega(n^2)$. Here we derive an independent
closed form for the same Gaussian moment. The formula refines this picture:
every fixed positive value of $k/n^2$ gives an inverse polynomial ratio, the
quadratic scale is a smooth Bessel crossover, and the order boundary occurs at
$k\asymp n^2/\log n$. The interferometer mode count enters only the separate
finite Haar hiding step.

Let $X=(x_1,\ldots,x_{2n})$ have independent standard circular complex
Gaussian entries, $\E|X_{ai}|^2=1$. The transpose in
$X^{\mathsf T}X$ is not a conjugate transpose. Define
\begin{equation}
 \begin{gathered}
 H_{k,n}=\haf(X^{\mathsf T}X),\qquad Q_{k,n}=|H_{k,n}|^2,\\[-2pt]
 M_1=\E Q_{k,n},\qquad M_2=\E Q_{k,n}^2,\qquad
 \R_{k,n}=\frac{M_1^2}{M_2}.
 \end{gathered}
 \label{eq:model}
\end{equation}
We use unit variance Gaussian entries. Replacing $X$ by $X/\sqrt m$, as in
the variance $1/m$ Gaussian surrogate for a selected Haar submatrix with $m$
interferometer modes, gives
$\haf[(X/\sqrt m)^{\mathsf T}(X/\sqrt m)]=m^{-n}H_{k,n}$, so the scale cancels
from $\R_{k,n}$. This rescaling is not an equality between finite Haar and
independent Gaussian ensembles; distributional approximation and moment
transfer require separate hypotheses. Throughout, $\log$ denotes the natural
logarithm. For a sequence $k_n\in\mathbb N_{>0}$, we call
$(\R_{k_n,n})_{n\ge1}$ weakly anticoncentrated if there exist $C>0$ and
$d\in\mathbb N$ such that $\R_{k_n,n}\ge Cn^{-d}$ for all sufficiently large
$n$. Indeed, the inequality of Paley and Zygmund gives
$\mathbb{P}\!\left(Q_{k,n}\ge\theta\E Q_{k,n}\right)\ge
(1-\theta)^2\R_{k,n}$ for $0<\theta<1$. This is the weak second moment
criterion used here; it is not a high probability lower tail statement.

\textit{Exact moment theorem.}
For integers $k,n\ge1$, write
\begin{equation}
 P_{k,n}=\prod_{q=0}^{n-1}(k+2q),\qquad
 F_{k,n}=\sum_{j=0}^{n}\binom{n}{j}^{\!2}
              \frac{(1/2)_j}{(k/2)_j},
 \label{eq:PF}
\end{equation}
where $(a)_j$ is the rising factorial. Then the first two moments of the
output weight are
\begin{equation}
 M_1=(2n-1)!!\,P_{k,n},\qquad
 M_2=(2n)!\,\bigl(P_{k,n}\bigr)^2F_{k,n},
 \label{eq:exact-moments}
\end{equation}
and therefore
\begin{equation}
 \R_{k,n}=\frac{C_n}{F_{k,n}},\qquad
 C_n=4^{-n}\binom{2n}{n}.
 \label{eq:exact-ratio}
\end{equation}
The factor $F_{k,n}$ in Eq.~\eqref{eq:PF} is a terminating generalized
hypergeometric polynomial, given explicitly in End Matter. The finite
polynomial can be evaluated directly without numerical recursion. All $k$
dependence of the ratio
$\R_{k,n}$ is carried by $F_{k,n}$; the central binomial factor satisfies
$C_n\sim(\pi n)^{-1/2}$. At the rank one endpoint,
$F_{1,n}=\binom{2n}{n}$ and $\R_{1,n}=4^{-n}$, whereas
$F_{k_n,n}\to1$ whenever $k_n/n^2\to\infty$. The exact finite sum thus
interpolates between an
exponentially small ratio and the universal central binomial baseline without
requiring numerical recursion.

\begin{figure*}[t]
 \centering
 \includegraphics[width=0.86\textwidth]{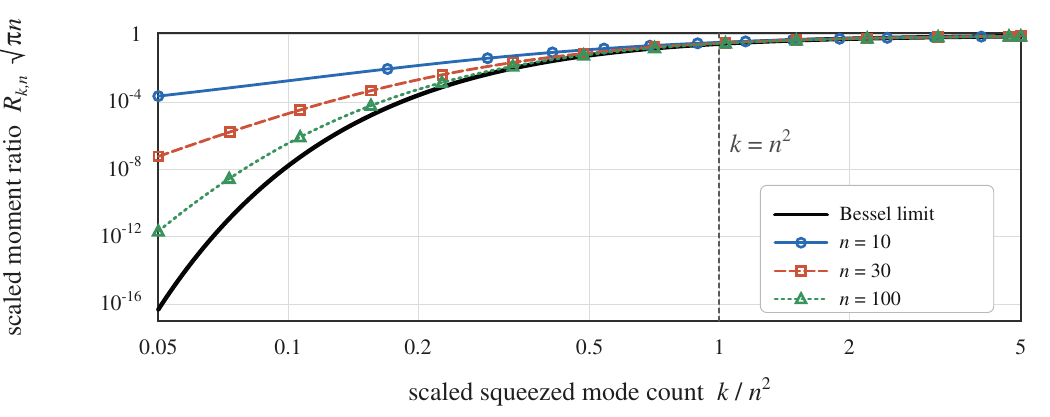}
 \caption{Exact finite sum ratio from Eq.~\eqref{eq:exact-ratio} for
 $n=10,30,100$, together with the Bessel limit in
 Eq.~\eqref{eq:bessel-crossover}. Both axes are logarithmic; symbols show
 exact finite $n$ evaluations, not simulated data, and the vertical line marks
 $k=n^2$. For every
 fixed $k/n^2>0$, the scaled ratio approaches a positive limit, so the
 quadratic scale is a crossover within the weak regime.}
 \label{fig:crossover}
\end{figure*}

\textit{Proof mechanism.}
Using the auxiliary field identity in End Matter,
Eq.~\eqref{eq:auxiliary}, each hafnian becomes an auxiliary Gaussian
expectation of a product of linear forms;
the full derivation begins at Eq.~(S8) of Supplemental
Material included below. For the fourth moment, the exact single column
identities in Eqs.~(S11) and (S12) leave only two Wick contractions.
Independence of the $2n$ columns then collapses the matching enumeration to
\begin{equation}
 M_2=\E\!\left[(g_1\!\cdot h_1)(g_2\!\cdot h_2)
             +(g_1\!\cdot h_2)(g_2\!\cdot h_1)\right]^{2n},
 \label{eq:rank-two}
\end{equation}
where the four vectors are independent $N_k(0,I_k)$. Conditional on
$g_1,g_2$, this is a Gaussian moment of rank at most two.
Supplemental Eqs.~(S13) through (S21) give the rank two diagonalization, the
$k\ge2$ almost sure qualification, the separate $k=1$ integral, and the
radial and angular calculation. In particular, radial integration produces
$\bigl(P_{k,n}\bigr)^2$, while the angle between two independent directions
has moments $(1/2)_j/(k/2)_j$, producing $F_{k,n}$.

This reduction is the conceptual reason a closed form exists. Directly
expanding four hafnians produces a large family of overlapping perfect
matchings, whose graph types proliferate with $n$. The auxiliary field sums
those matchings before the random matrix expectation is taken. The two
allowed complex Wick contractions then leave only the two dimensional span
of $g_1$ and $g_2$. Rotational invariance separates their radii from their
angle, so the matching expansion reduces to moments of one beta distributed
squared directional correlation. In particular, no saddle point or large $n$
approximation enters Eqs.~\eqref{eq:exact-moments} and
\eqref{eq:exact-ratio}.

\textit{Two scales.}
Set $B(y)=e^yI_0(y)$, where $I_0$ is the modified Bessel function of the
first kind of order zero. A coefficientwise comparison of the exact finite
sum with the Bessel series gives, for all $k,n\ge1$,
\begin{equation}
 1\le F_{k,n}\le B(n^2/k)\le e^{2n^2/k},\qquad
 \R_{k,n}\ge C_n e^{-2n^2/k}.
 \label{eq:finite-bounds}
\end{equation}
At the quadratic scale, dominated convergence in Eq.~\eqref{eq:PF} yields
\begin{equation}
 \frac{k_n}{n^2}\to c>0
 \quad\Longrightarrow\quad
 \begin{cases}
 F_{k_n,n}\to e^{1/c}I_0(1/c),\\[2pt]
 \R_{k_n,n}\sqrt{\pi n}\to[e^{1/c}I_0(1/c)]^{-1}.
 \end{cases}
 \label{eq:bessel-crossover}
\end{equation}
This conclusion refines the picture suggested by the earlier $k=O(n^2)$
conjecture: every fixed $c>0$ already yields an inverse polynomial ratio. The
exact finite $n$ approach to this limit is shown in Fig.~\ref{fig:crossover}.
Write $y=1/c$ and let $a_{k,n,j}$ denote the $j$th summand in
Eq.~\eqref{eq:PF}, extended by zero for $j>n$. For each fixed $j$,
$a_{k_n,n,j}\to b_j(y)=(1/2)_j(2y)^j/(j!)^2$. For all sufficiently large
$n$, $0\le a_{k_n,n,j}\mathbf 1_{\{j\le n\}}\le b_j(y+1)$, and
$\sum_j b_j(y+1)<\infty$. Dominated convergence for series therefore yields
Eq.~\eqref{eq:bessel-crossover}. This explains why
the quadratic scale changes only a finite prefactor multiplying
$C_n\sim(\pi n)^{-1/2}$.

The qualitative change occurs earlier. The lower bound in
Eq.~\eqref{eq:finite-bounds} proves weak anticoncentration whenever
$n^2/k_n=O(\log n)$. Conversely, a one term witness defeats every eventual
polynomial lower bound in the converse regime. For every eventually positive
integer sequence $k_n$,
\begin{equation}
 \begin{aligned}
 n^2/k_n=O(\log n)
   &\Longrightarrow
   \mathrm{WeakAC}\!\left((\R_{k_n,n})_{n\ge1}\right),\\
 k_n\log n/n^2\to0
   &\Longrightarrow
   \neg\mathrm{WeakAC}\!\left((\R_{k_n,n})_{n\ge1}\right).
 \end{aligned}
 \label{eq:threshold}
\end{equation}
Equivalently, $k_n=\Omega(n^2/\log n)$ is sufficient, whereas
$k_n=o(n^2/\log n)$ forces failure of weak anticoncentration. Here and in the
title, threshold means a boundary in scaling order; no universal critical
constant is asserted within the boundary window. The logarithm appears because
weak anticoncentration tolerates polynomial decay whereas $F_{k,n}$ can grow
exponentially in $n^2/k$.
If $k_n\log n/n^2\to\beta>0$, then eventually
$\R_{k_n,n}\ge[2n^{\lceil4/\beta\rceil+1}]^{-1}$, and hence weak
anticoncentration holds. This exponent is a certificate, not a sharp exponent.
Conversely, if $k_n$ is smaller than $n^2/\log n$ by an unbounded factor, the
single term witness in End Matter defeats every proposed polynomial lower
bound. The logarithmic scale is therefore an order boundary, whereas
$k\asymp n^2$ is an analytic crossover inside the weak regime.

The exact ratio also determines the averaged ideal Gaussian surrogate LXEB
reference scale through
$\E F_{\mathrm{XEB}}^{\mathrm{ideal}}=\R_{k,n}^{-1}-1$
\cite{EhrenbergPRA2025}. The closed form gives its asymptotic normalization
throughout the quadratic window and replaces a graph recursion by a directly
evaluable finite polynomial. For broader LXEB context, see
Ref.~\cite{MartinezCifuentes2024}; a concurrent finite Haar framework likewise
distinguishes the second moment criterion from full GBS anticoncentration
\cite{KolarovszkiEtAl2026}.

We have obtained an exact finite formula for the first two probability moments
of a Gaussian Gram hafnian. The terminating generalized hypergeometric
function ${}_3F_2$ encodes the finite $k,n$ dependence, while its modified
Bessel limit separates the quadratic crossover from the logarithmically
smaller order boundary for weak anticoncentration. Thus the independent
Gaussian normalized second moment step analyzed here is now exact. Extending
the result to
finite dimensional Haar interferometers requires moment sensitive hiding;
distributional hiding alone does not transfer the unbounded moments
\cite{ShouMillerGalitski2025}. Full small ball anticoncentration
requires additional lower tail information. The rank two reduction may also
help analyze correlated Gaussian matrix functions whose matching expansions
obscure low dimensional structure.

\section*{Acknowledgments}
\textit{AI use disclosure.}
OpenAI Codex, using GPT 5.6 Sol in Ultra mode,
assisted with literature organization, LaTeX restructuring, language editing,
formalization in Lean of all sixty four numbered mathematical equations in the
Letter and Supplemental Material, and execution of Lean kernel checks. The
author directed the work,
checked the mathematical
statements and citations against the source material, reviewed the Lean
formulations, build results, and axiom reports, and takes full responsibility
for the manuscript.

\paragraph*{Data and software availability.}
No experimental data were generated. The Lean archive is titled
\emph{Lean Verification for ``Exact Moments of Gaussian Gram Hafnians Reveal
an \(n^2/\log n\) Threshold for Weak Anticoncentration''}. It is licensed
under the GNU General Public License v3.0 only and is publicly available on
Zenodo~\cite{ZhaoLeanVerification2026} at
\href{https://doi.org/10.5281/zenodo.21959946}{10.5281/zenodo.21959946}.

\nocite{deMouraUllrich2021,Mathlib2020}
\bibliographystyle{unsrt}
\bibliography{references_prl}

\clearpage
\section*{End Matter}

\paragraph{Appendix A: Exact Gaussian reduction.}
Let $g\sim N_k(0,I_k)$ be real and independent of $X$. Wick's rule gives the
deterministic auxiliary field identity
\begin{equation}
 \haf(X^{\mathsf T}X)=\E_g\prod_{i=1}^{2n}g^{\mathsf T}x_i.
 \label{eq:auxiliary}
\end{equation}
Applying Eq.~\eqref{eq:auxiliary} to two copies and two conjugate copies,
then integrating one complex Gaussian column at a time, proves
Eq.~\eqref{eq:rank-two}. All exchanges of expectation are justified by
absolute integrability of Gaussian polynomials.

For $k\ge2$, define
\[
 \begin{gathered}
 \mathcal Q=(g_1\!\cdot h_1)(g_2\!\cdot h_2)
 +(g_1\!\cdot h_2)(g_2\!\cdot h_1)=h_1^{\mathsf T}Th_2,\\
 T=g_1g_2^{\mathsf T}+g_2g_1^{\mathsf T}.
 \end{gathered}
\]
On the almost sure event that $g_1,g_2$ are linearly independent, put
$A=\|g_1\|$, $B=\|g_2\|$, and $r=(g_1\cdot g_2)/(AB)$. The two nonzero
eigenvalues of $T$ are $AB(r+1)$ and $AB(r-1)$, giving
\begin{equation}
 \E_{h_1,h_2}\mathcal Q^{2n}
 =(2n)!(A^2B^2)^n\sum_{j=0}^{n}\binom{n}{j}^{\!2}r^{2j}.
 \label{eq:conditional-rank-two}
\end{equation}
The integrated $k=1$ endpoint follows from a separate scalar calculation.
For $k\ge2$, $A^2$ and $B^2$ have the chi squared law $\chi_k^2$, while
$r^2\sim\operatorname{Beta}\!\left(1/2,(k-1)/2\right)$, and all three
variables are mutually independent. Consequently,
$\E A^{2n}=P_{k,n}$ and $\E r^{2j}=(1/2)_j/(k/2)_j$; substitution proves
Eqs.~\eqref{eq:exact-moments} and \eqref{eq:exact-ratio}. Algebraically, the
resulting finite factor is the terminating generalized hypergeometric function
\begin{equation}
 F_{k,n}={}_3F_2\!\left(
 \begin{matrix}-n,-n,1/2\\1,k/2\end{matrix};1\right),
 \label{eq:threeFtwo}
\end{equation}
where the notation denotes the terminating coefficient sum in
Eq.~\eqref{eq:PF}.

\paragraph{Appendix B: Bounds and asymptotic regimes.}
The comparison function used in Eq.~\eqref{eq:finite-bounds} has the exact
series representation
\begin{equation}
 B(y)=\sum_{j=0}^{\infty}\frac{(1/2)_j}{(j!)^2}(2y)^j
     =e^yI_0(y),\qquad y\ge0.
 \label{eq:bessel-majorant}
\end{equation}
Termwise domination of Eq.~\eqref{eq:PF} by this series proves the first
bound in Eq.~\eqref{eq:finite-bounds}; comparison with the exponential
series proves the second. If $n^2/k_n\to0$, this factor tends to one and
\begin{equation}
 \frac{\R_{k_n,n}}{C_n}\longrightarrow1,\qquad
 C_n\sqrt{\pi n}\longrightarrow1.
 \label{eq:superquadratic}
\end{equation}

For the converse regime, positivity lets one retain a single summand of
$F_{k,n}$. For $k\ge1$ and $j\le n$,
\begin{equation}
 2j(k+2j)\le(n+1-j)^2
 \quad\Longrightarrow\quad
 \R_{k,n}\le2^{-j}.
 \label{eq:one-term-certificate}
\end{equation}
When $k_n\log n/n^2\to0$, a logarithmically growing witness $j=j_n$
satisfies the premise eventually and defeats every eventual
inverse polynomial lower bound. This proves the converse half of
Eq.~\eqref{eq:threshold}.

\paragraph{Formal verification scope.}
The archived Lean 4 development
\cite{deMouraUllrich2021,Mathlib2020,ZhaoLeanVerification2026} provides an
explicit declaration or theorem specialization for every numbered mathematical
equation in the Letter and Supplemental Material; the complete equation to
declaration map is supplied in the repository and in Supplemental Material.
Every mathematical result asserted in this work within the Gaussian surrogate
scope is proved by a named Lean theorem and checked by the Lean kernel.
Definitions and notational
equations are checked by elaboration and mapped to their defining declarations.
Equation~\eqref{eq:conditional-rank-two} is an almost sure identity for
$k\ge2$, while the integrated $k=1$ moment is verified separately. Axiom
reports for all Letter and Supplemental equation endpoints list only
\texttt{propext},
\texttt{Classical.choice}, and \texttt{Quot.sound}; a source scan of the
audited Gaussian Gram hafnian tree found no \texttt{sorry}, \texttt{admit}, or
project specific \texttt{axiom}. The verification excludes finite Haar
transfer, full small ball anticoncentration, cited literature, figure
generation, and physical interpretation; none of these is a
mathematical result claimed by the paper.

\clearpage
\onecolumn
\appendix
\setcounter{section}{2}
\renewcommand{\theequation}{S\arabic{equation}}
\setcounter{equation}{0}
\section*{Appendices: Supplemental Material}

\section{Scope and conventions}

This supplement proves the finite moment formula, its finite bounds, the
quadratic Bessel crossover, the central binomial baseline, and both sides of
the logarithmic weak anticoncentration boundary. The model is the independent
complex Gaussian surrogate used in the collision free analysis of Gaussian
boson sampling (GBS) \cite{Hamilton2017}; see also
Refs.~\cite{EhrenbergPRL2025,EhrenbergPRA2025}. No finite Haar hiding statement
and no small ball theorem is asserted here. Throughout, $\log$ denotes the
natural logarithm.

The equation to declaration map below records the formal counterpart of every
numbered mathematical equation. Source comments beginning with
\texttt{LEAN:} identify the corresponding declarations.

Let a standard circular complex Gaussian be realized as
\begin{equation}
 Z=\frac{G_1+\mathrm iG_2}{\sqrt2},
 \qquad G_1,G_2\overset{\mathrm{IID}}{\sim}\N(0,1),
 \qquad \E|Z|^2=1,
 \label{eq:S-circular}
\end{equation}
and let $X=(x_1,\ldots,x_{2n})\in\C^{k\times 2n}$ have independent
entries with this law.  The transpose Gram matrix is $X^{\mathsf T}X$; no
complex conjugation is taken inside this matrix.  For a symmetric
$2n\times2n$ matrix $A$, we use
\begin{equation}
 \haf(A)=\sum_{M\in\mathfrak M_{2n}}
        \prod_{\{a,b\}\in M} A_{ab},
 \qquad H_{k,n}=\haf(X^{\mathsf T}X),
 \label{eq:S-hafnian}
\end{equation}
where $\mathfrak M_{2n}$ is the set of perfect matchings. Throughout the
statements presented in the paper, $k,n\ge1$; the algebraic formulas also
extend to $n=0$ for every positive integer $k$.

Define
\begin{equation}
 M_1(k,n)=\E|H_{k,n}|^2,
 \qquad M_2(k,n)=\E|H_{k,n}|^4,
 \qquad \R_{k,n}=\frac{M_1(k,n)^2}{M_2(k,n)}.
 \label{eq:S-moments}
\end{equation}
The deterministic quantities appearing in the answer are
\begin{align}
 D_n&=(2n-1)!!=\prod_{q=0}^{n-1}(2q+1),
 &P_{k,n}&=\prod_{q=0}^{n-1}(k+2q),
 \label{eq:S-DP}\\
 F_{k,n}&=\sum_{j=0}^{n}a_{k,n,j},
 &a_{k,n,j}&=\binom nj^2\frac{(1/2)_j}{(k/2)_j},
 \label{eq:S-F}\\
 C_n&=4^{-n}\binom{2n}{n},
 &r_{k,n}&=\frac{C_n}{F_{k,n}}.
 \label{eq:S-Cr}
\end{align}
Here $(a)_j=a(a+1)\cdots(a+j-1)$ and $(a)_0=1$.  The finite product used in
Lean for the Pochhammer quotient is
\begin{equation}
 \frac{(1/2)_j}{(k/2)_j}
   =\prod_{\ell=0}^{j-1}\frac{2\ell+1}{k+2\ell}.
 \label{eq:S-poch-product}
\end{equation}
This product form makes positivity and all denominator conditions explicit.

\section{Auxiliary field compression and justified Fubini interchange}
\label{sec:auxiliary}

Let $g\sim\N_k(0,I_k)$ be real and independent of $X$.  For each fixed
complex matrix $X$, the auxiliary field identity is
\begin{equation}
 H_{k,n}=\E_g\prod_{i=1}^{2n}g^{\mathsf T}x_i.
 \label{eq:S-auxiliary}
\end{equation}
To see the combinatorics, expand the product by choosing a row color for
each column.  A coordinate of $g$ contributes zero unless its multiplicity
is even; for multiplicity $2r$, its expectation is $(2r-1)!!$.  Regrouping
the compatible coordinate pairings gives precisely the perfect matching sum
in Eq.~\eqref{eq:S-hafnian}. This finite regrouping is a bijection between
the compatible coordinate pairings and the colored perfect matchings.

The moment proof uses two and four independent auxiliary fields.  All
integrands are finite sums of Gaussian monomials.  Their absolute
integrability is proved term by term, and hence Fubini's and Tonelli's
theorems give
\begin{align}
 M_1(k,n)
 &=\E_{g,h}(g\mathbin\cdot h)^{2n},
 \label{eq:S-M1-reduction}\\
 M_2(k,n)
 &=\E_{g_1,g_2,h_1,h_2}
   \left[(g_1\mathbin\cdot h_1)(g_2\mathbin\cdot h_2)
        +(g_1\mathbin\cdot h_2)(g_2\mathbin\cdot h_1)\right]^{2n}.
 \label{eq:S-M2-reduction}
\end{align}
Indeed, for one circular complex Gaussian column $x$, the two identities
used after interchanging expectations are
\begin{align}
 \E_x[(g^{\mathsf T}x)(h^{\mathsf T}\bar x)]
 &=g\mathbin\cdot h,
 \label{eq:S-column-two}\\
 \E_x\!\left[\prod_{a=1}^{2}(g_a^{\mathsf T}x)
                   \prod_{b=1}^{2}(h_b^{\mathsf T}\bar x)\right]
 &=(g_1\mathbin\cdot h_1)(g_2\mathbin\cdot h_2)
  +(g_1\mathbin\cdot h_2)(g_2\mathbin\cdot h_1).
 \label{eq:S-column-four}
\end{align}
The $2n$ independent columns raise each quantity on the right to the power
$2n$. Thus Eqs.~\eqref{eq:S-M1-reduction} and
\eqref{eq:S-M2-reduction} are exact equalities of finite dimensional Gaussian
integrals.

\section{Rank two conditional moment and the exact finite formula}
\label{sec:ranktwo}

Put
\begin{equation}
 Q(g_1,g_2;h_1,h_2)
 =(g_1\mathbin\cdot h_1)(g_2\mathbin\cdot h_2)
 +(g_1\mathbin\cdot h_2)(g_2\mathbin\cdot h_1).
 \label{eq:S-Q}
\end{equation}
For fixed $g_1,g_2$, $Q$ depends on each $h_a$ only through the span of
$g_1,g_2$.  The symmetric rank two operator
\begin{equation}
 T=g_1g_2^{\mathsf T}+g_2g_1^{\mathsf T}
 \quad\text{has, for linearly independent }g_1,g_2,\text{ eigenvalues}\quad
 \lambda_{\pm}=g_1\mathbin\cdot g_2\pm\|g_1\|\,\|g_2\|.
 \label{eq:S-eigenvalues}
\end{equation}
Diagonalizing $T$ inside this span of dimension at most two and evaluating the
two independent one dimensional Gaussian coordinates gives
\begin{equation}
 \E_{h_1,h_2}Q^{2n}
 =(2n)!\sum_{j=0}^{n}\binom nj^2
    (\|g_1\|^2\|g_2\|^2)^{n-j}
    (g_1\mathbin\cdot g_2)^{2j}.
 \label{eq:S-conditional}
\end{equation}

Equation~\eqref{eq:S-conditional} is first proved when $g_1,g_2$ are linearly
independent. For $k\ge2$,
\begin{equation}
 \mathbb{P}\!\left(g_1,g_2\ \text{are linearly independent}\right)=1.
 \label{eq:S-ae-LI}
\end{equation}
Consequently the conditional identity can be substituted under the outer
integral without changing its value. The endpoint $k=1$, where linear
independence is impossible, follows from a separate one dimensional
calculation. Hence the final moment theorem covers every positive integer
$k$.

For the outer integral, write
\begin{equation}
 A^2=\|g_1\|^2,\qquad B^2=\|g_2\|^2,\qquad
 U=\frac{(g_1\mathbin\cdot g_2)^2}{\|g_1\|^2\|g_2\|^2},
 \label{eq:S-radial-angle}
\end{equation}
with $U$ defined arbitrarily on the null event $AB=0$. In dimension $k\ge2$,
the exact joint law is
\begin{equation}
 A^2\sim\chi_k^2,\qquad
 U\sim\operatorname{Beta}\!\left(\frac12,\frac{k-1}{2}\right),\qquad
 B^2\sim\chi_k^2,\qquad A^2,U,B^2\ \text{mutually independent},
 \label{eq:S-product-law}
\end{equation}
Therefore,
\begin{align}
 \E A^{2n}=\E B^{2n}&=P_{k,n},
 &\E U^j&=\frac{(1/2)_j}{(k/2)_j},
 \label{eq:S-radial-moments}\\
 \E\!\left[(A^2B^2)^{n-j}(g_1\mathbin\cdot g_2)^{2j}\right]
 &=\bigl(P_{k,n}\bigr)^2\frac{(1/2)_j}{(k/2)_j}.
 \label{eq:S-outer-monomial}
\end{align}
Substitution in Eq.~\eqref{eq:S-conditional} proves the fourth moment
formula.  The corresponding rank one integral in
Eq.~\eqref{eq:S-M1-reduction} gives the first moment:
\begin{equation}
 \boxed{
 M_1(k,n)=D_nP_{k,n},\qquad
 M_2(k,n)=(2n)!\bigl(P_{k,n}\bigr)^2F_{k,n}.}
 \label{eq:S-exact-moments}
\end{equation}
The elementary identity $D_n^2/(2n)!=C_n$ cancels the common radial factor,
so
\begin{equation}
 \boxed{\R_{k,n}=r_{k,n}=\frac{C_n}{F_{k,n}}
 =\frac{4^{-n}\binom{2n}{n}}
 {\displaystyle\sum_{j=0}^{n}\binom nj^2(1/2)_j/(k/2)_j}.}
 \label{eq:S-exact-ratio}
\end{equation}

Finally, using $(-n)_j/j!=(-1)^j\binom nj$ term by term gives the exact
terminating hypergeometric representation
\begin{equation}
 \boxed{F_{k,n}={}_3F_2\!\left(
 \begin{matrix}-n,-n,1/2\\1,k/2\end{matrix};1\right).}
 \label{eq:S-threeFtwo}
\end{equation}
Because the coefficient sum terminates, Eq.~\eqref{eq:S-threeFtwo} requires
no analytic continuation.

\section{Finite Bessel and exponential bounds}
\label{sec:finite}

Set $y=n^2/k$ and define the following series, which converge everywhere,
\begin{align}
 B(y)&=\sum_{j=0}^{\infty}b_j(y),
 &b_j(y)&=\frac{(1/2)_j(2y)^j}{(j!)^2},
 \label{eq:S-B}\\
 I_0(y)&=\sum_{r=0}^{\infty}\frac{y^{2r}}{4^r(r!)^2}.
 \label{eq:S-I0}
\end{align}
For $j\le n$, the elementary estimates
\begin{equation}
 \binom nj\le\frac{n^j}{j!},\qquad
 \frac{(1/2)_j}{(k/2)_j}\le (1/2)_j\left(\frac2k\right)^j
 \quad\Longrightarrow\quad a_{k,n,j}\le b_j(y)
 \label{eq:S-term-majorant}
\end{equation}
give the first nonasymptotic majorant.  An absolutely convergent Cauchy
product proves the exact identity
\begin{equation}
 \boxed{B(y)=e^yI_0(y)\quad\text{for every real }y.}
 \label{eq:S-B-I0}
\end{equation}
The finite coefficient identity behind this Cauchy product is obtained by
equating the middle coefficients of
\begin{equation}
 (1+2z+z^2)^\ell=(1+z)^{2\ell}.
 \label{eq:S-coefficient-identity}
\end{equation}
Because $(1/2)_j\le j!$ and all terms are nonnegative for $y\ge0$, we obtain
\begin{equation}
 \boxed{
 1\le F_{k,n}\le B(n^2/k)=e^{n^2/k}I_0(n^2/k)
 \le e^{2n^2/k},}
 \label{eq:S-finite-chain}
\end{equation}
and hence
\begin{equation}
 \boxed{
 \frac{C_n}{B(n^2/k)}\le\R_{k,n},\qquad
 C_ne^{-2n^2/k}\le\R_{k,n}\le C_n.}
 \label{eq:S-ratio-chain}
\end{equation}
These bounds are uniform over all positive integral $k$ and all $n$.

The converse at the logarithmic boundary uses a lower bound on one summand.
For $j\le n$,
\begin{equation}
 a_{k,n,j}\ge
 \frac{(n+1-j)^{2j}}{j!\,(k+2j)^j}.
 \label{eq:S-one-term-lower}
\end{equation}
Since $j!\le j^j$, the transparent geometric certificate
\begin{equation}
 \boxed{
 2j(k+2j)\le(n+1-j)^2
 \quad\Longrightarrow\quad
 \R_{k,n}\le2^{-j}.}
 \label{eq:S-geometric-certificate}
\end{equation}
Indeed, the hypothesis makes the quantity on the right in
Eq.~\eqref{eq:S-one-term-lower} at least $2^j$, whereas $C_n\le1$.

\section{Quadratic Bessel crossover and the superquadratic baseline}
\label{sec:quadratic}

Let $k_n$ take natural values, suppose $k_n\to\infty$, and assume
\begin{equation}
 \frac{n^2}{k_n}\longrightarrow y\in[0,\infty).
 \label{eq:S-y-limit}
\end{equation}
Then
\begin{equation}
 \boxed{F_{k_n,n}\longrightarrow B(y).}
 \label{eq:S-general-B-limit}
\end{equation}
For completeness, extend $a_{k,n,j}$ by zero for $j>n$. For $0\le j<n$,
the exact recurrence is
\begin{equation}
 \frac{a_{k,n,j+1}}{a_{k,n,j}}
 =\frac{(n-j)^2(2j+1)}{(j+1)^2(k+2j)}.
 \label{eq:S-recurrence}
\end{equation}
For every fixed $j$, the condition $j<n$ holds eventually. Under
Eq.~\eqref{eq:S-y-limit}, this ratio converges to
$y(2j+1)/(j+1)^2$, which is exactly $b_{j+1}(y)/b_j(y)$.  Since
$a_{k,n,0}=b_0(y)=1$, induction proves pointwise convergence of every fixed
summand.  Eventually $n^2/k_n\le y+1$, and
Eq.~\eqref{eq:S-term-majorant} gives the summable domination
\begin{equation}
 0\le a_{k_n,n,j}\mathbf 1_{\{j\le n\}}\le b_j(y+1).
 \label{eq:S-Tannery-bound}
\end{equation}
Tannery's theorem (dominated convergence for series) now proves
Eq.~\eqref{eq:S-general-B-limit}.

If instead $k_n/n^2\to c>0$, then $k_n\to\infty$ follows automatically and
the reciprocal limit gives $y=1/c$. Therefore,
\begin{equation}
 \boxed{F_{k_n,n}\longrightarrow e^{1/c}I_0(1/c).}
 \label{eq:S-quadratic-F}
\end{equation}
Stirling's formula yields
\begin{equation}
 C_n\sqrt{\pi n}\longrightarrow1.
 \label{eq:S-central-Stirling}
\end{equation}
Combining Eqs.~\eqref{eq:S-exact-ratio}, \eqref{eq:S-quadratic-F}, and
\eqref{eq:S-central-Stirling} gives the Gaussian crossover
\begin{equation}
 \boxed{
 \R_{k_n,n}\sqrt{\pi n}\longrightarrow
 \frac{1}{e^{1/c}I_0(1/c)}.}
 \label{eq:S-quadratic-R}
\end{equation}
Thus every fixed positive quadratic aspect ratio lies inside the weak
second moment anticoncentration regime.

When $k_n$ is eventually positive and $n^2/k_n\to0$, the elementary squeeze
in Eq.~\eqref{eq:S-finite-chain} already gives
\begin{equation}
 \boxed{F_{k_n,n}\longrightarrow1,\qquad
 \frac{\R_{k_n,n}}{C_n}\longrightarrow1.}
 \label{eq:S-superquadratic}
\end{equation}
Together with Eq.~\eqref{eq:S-central-Stirling}, this is the
central binomial baseline quoted in the Letter.

\section{Logarithmic weak anticoncentration boundary}
\label{sec:logboundary}

For a real sequence $q_n$, weak anticoncentration means
\begin{equation}
 \operatorname{WeakAC}(q)\quad\Longleftrightarrow\quad
 \exists C>0\ \exists d\in\mathbb N\ \text{such that for all sufficiently large }n,
 \quad
 \frac{C}{n^d}\le q_n.
 \label{eq:S-weak-def}
\end{equation}
An elementary central binomial inequality gives $C_n\ge1/(2n)$ for $n>0$.
Combining it with the exponential bound in
Eq.~\eqref{eq:S-finite-chain} proves the pointwise implication
\begin{equation}
 2\frac{n^2}{k}\le d\log n
 \quad\Longrightarrow\quad
 \boxed{\R_{k,n}\ge\frac{1}{2n^{d+1}}}
 \qquad(k,n>0,\ d\in\mathbb N).
 \label{eq:S-log-sufficient-pointwise}
\end{equation}
Consequently, if $k_n$ is eventually positive and the hypothesis on the left
of Eq.~\eqref{eq:S-log-sufficient-pointwise} holds eventually for one fixed
$d$, then
\begin{equation}
 \boxed{\operatorname{WeakAC}\!\left((\R_{k_n,n})_{n\ge1}\right).}
 \label{eq:S-log-sufficient}
\end{equation}

A convenient corollary for a positive window is also formalized.  If $k_n$ is
eventually positive and
\begin{equation}
 \frac{k_n\log n}{n^2}\longrightarrow\beta>0,
 \label{eq:S-beta-positive}
\end{equation}
then eventually
\begin{equation}
 \boxed{
 \R_{k_n,n}\ge
 \frac{1}{2n^{\lceil4/\beta\rceil+1}},}
 \label{eq:S-beta-lower}
\end{equation}
and hence Eq.~\eqref{eq:S-log-sufficient} holds.  The exponent in
Eq.~\eqref{eq:S-beta-lower} is a proved certificate, not a claim of sharpness.

For the converse, assume $k_n$ is eventually positive and
\begin{equation}
 \frac{k_n\log n}{n^2}\longrightarrow0.
 \label{eq:S-o-regime}
\end{equation}
Given any proposed polynomial degree $d$, choose
\begin{equation}
 j_n=\left\lceil\frac{d+1}{\log2}\log n\right\rceil,
 \qquad n^{d+1}\le2^{j_n}.
 \label{eq:S-log-witness}
\end{equation}
Then $j_n/n\to0$.  Moreover, Eq.~\eqref{eq:S-o-regime} implies
\begin{equation}
 \frac{2j_n(k_n+2j_n)}{n^2}\longrightarrow0.
 \label{eq:S-certificate-limit}
\end{equation}
Thus, eventually, $j_n\le n$ and the hypothesis in
Eq.~\eqref{eq:S-geometric-certificate} holds.  It follows that
$\R_{k_n,n}\le2^{-j_n}\le n^{-(d+1)}$.  This contradicts any eventual lower
bound $C/n^d$, since $1/n<C$ eventually.  Therefore
\begin{equation}
 \boxed{
 \frac{k_n\log n}{n^2}\longrightarrow0
 \quad\Longrightarrow\quad
 \neg\operatorname{WeakAC}\!\left((\R_{k_n,n})_{n\ge1}\right).}
 \label{eq:S-log-converse}
\end{equation}
At the exact rank one endpoint the formula simplifies further:
\begin{equation}
 F_{1,n}=\binom{2n}{n},\qquad
 \boxed{\R_{1,n}=4^{-n}},
 \label{eq:S-rank-one}
\end{equation}
so weak anticoncentration fails exponentially.

Equations~\eqref{eq:S-log-sufficient} and \eqref{eq:S-log-converse} are the
proved order statement.  They do not assert a sharp leading constant at
every sequence inside the boundary window.

\section{Relation to finite dimensional Haar interferometers}
\label{sec:haar-scope}

The exact results above concern the independent complex Gaussian surrogate.
They do not by themselves establish the corresponding moment statement for a
finite dimensional Haar interferometer. Theorem 1.5 of
Ref.~\cite{ShouMillerGalitski2025} assumes $1\le2n\le k<m$ and
$(2n)k=o(m)$, where $m$ is the number of interferometer modes. At the weak
anticoncentration order boundary $k\asymp n^2/\log n$, the latter condition
requires $m\gg n^3/\log n$.

Moreover, total variation control of the selected submatrix law does not by
itself transfer the unbounded hafnian moments entering $\R_{k,n}$. Such a
transfer requires a moment sensitive estimate. The finite Gaussian moment
theorem proved here and the finite Haar hiding problem are therefore separate
statements, and no finite Haar moment transfer is claimed in this work.

\section{Equation to Lean crosswalk}
\label{sec:crosswalk}

The table lists the exact declaration that verifies each result in the paper.
Paths are relative to the GramHafnian source directory.  Definitions
and purely notational equations are mapped to their defining declarations.

\begingroup
\footnotesize
\setlength{\LTleft}{0pt}
\setlength{\LTright}{0pt}
\renewcommand{\arraystretch}{1.14}
\begin{longtable}{@{}>{\raggedright\arraybackslash}p{0.15\textwidth}
 >{\raggedright\arraybackslash}p{0.80\textwidth}@{}}
\toprule
Paper item & Exact Lean declaration and source file \\
\midrule
\endfirsthead
\toprule
Paper item & Exact Lean declaration and source file \\
\midrule
\endhead
\eqref{eq:S-circular} & \LeanDecl{circularGaussianCoordinate};
\LeanDecl{circularGaussian}; \LeanDecl{integral_mul_conj_circularGaussian}
(\emph{CircularGaussianMoments.lean}) \\
\eqref{eq:S-hafnian} & \LeanDecl{hafnian}; \LeanDecl{gramHafnian};
\LeanDecl{gramHafnian_eq_sum_coloredMatchings} (\emph{Hafnian.lean}) \\
\eqref{eq:S-moments} & \LeanDecl{actualGramFirstMomentReal};
\LeanDecl{actualGramFourthMomentReal}; \LeanDecl{actualGramSecondMomentRatio}
(\emph{ActualGramMoments.lean}; \emph{ExactGramMoments.lean}) \\
\eqref{eq:S-DP}, \eqref{eq:S-F}, and \eqref{eq:S-Cr} & \LeanDecl{oddPairingNat};
\LeanDecl{dimensionProduct}; \LeanDecl{gramHafnianFactor};
\LeanDecl{gramHafnianFactorTerm}; \LeanDecl{gramHafnianFactor_eq_sum_terms};
\LeanDecl{gramHafnianFactorTerm_eq_pochhammer};
\LeanDecl{centralBaseline}; \LeanDecl{gramSecondMomentRatio_eq_factor}
(\emph{FiniteMomentAlgebra.lean}; \emph{FiniteSum.lean};
\emph{PRLEquations.lean}) \\
\eqref{eq:S-poch-product} & \LeanDecl{pochhammerRatio};
\LeanDecl{half_rising_div_dimension_rising}
(\emph{FiniteSum.lean}; \emph{FiniteHypergeometric.lean}) \\
\eqref{eq:S-auxiliary} & \LeanDecl{integral_complexAuxiliaryFieldProduct_eq_gramHafnian}
(\emph{WickRegrouping.lean}) \\
\eqref{eq:S-M1-reduction} and \eqref{eq:S-M2-reduction} &
\LeanDecl{actualGramFirstMomentReal_eq_realGaussianBilinearIntegral};
\LeanDecl{actualGramFourthMomentReal_eq_realGaussianRankTwoIntegral}
(\emph{ActualGramMoments.lean}) \\
\eqref{eq:S-column-two} and \eqref{eq:S-column-four} &
\LeanDecl{integral_columnSecondIntegrand_circularGaussianVector};
\LeanDecl{integral_columnFourthIntegrand_circularGaussianVector}
(\emph{CircularGaussianSecondWick.lean}; \emph{CircularGaussianVectorWick.lean}) \\
\eqref{eq:S-Q} & \LeanDecl{innerRankTwoBilinear}
(\emph{RankTwoOrthogonalInvariance.lean}) \\
\eqref{eq:S-eigenvalues} & \LeanDecl{rankTwoAction};
\LeanDecl{rankTwoAction_normalize_eigenPlus};
\LeanDecl{rankTwoAction_normalize_eigenMinus};
\LeanDecl{rankTwoEigenPlus_norm_ne_zero_of_linearIndependent};
\LeanDecl{rankTwoEigenMinus_norm_ne_zero_of_linearIndependent}
(\emph{RankTwoEigenReduction.lean}; \emph{RankTwoConditionalMoment.lean}) \\
\eqref{eq:S-conditional} &
\LeanDecl{integral_innerRankTwoBilinear_pow_two_mul_of_linearIndependent}
(\emph{RankTwoConditionalMoment.lean}) \\
\eqref{eq:S-ae-LI} & \LeanDecl{ae_linearIndependent_gaussianPair}
(\emph{GaussianPairLinearIndependence.lean}) \\
\eqref{eq:S-radial-angle} and \eqref{eq:S-product-law} &
\LeanDecl{gaussianPairRadialAngleCoordinates};
\LeanDecl{map_gaussianPairRadialAngleCoordinates_gaussianProduct}
(\emph{RankTwoOuterBeta.lean}) \\
\eqref{eq:S-radial-moments} and \eqref{eq:S-outer-monomial} &
\LeanDecl{integral_pow_gammaMeasure_half_eq_dimensionProduct};
\LeanDecl{integral_pow_squaredAngleBeta_eq_pochhammerRatio};
\LeanDecl{integral_innerMonomial_gaussianProduct}
(\emph{RankOneGaussianBilinear.lean}; \emph{RankTwoOuterBeta.lean}) \\
\eqref{eq:S-exact-moments} &
\LeanDecl{prlEquation_exactMoments} (\emph{PRLEquations.lean}) \\
\eqref{eq:S-exact-ratio} &
\LeanDecl{prlEquation_exactRatio} (\emph{PRLEquations.lean}) \\
\eqref{eq:S-threeFtwo} &
\LeanDecl{prlEquation_threeFtwo} (\emph{PRLEquations.lean}) \\
\eqref{eq:S-B} and \eqref{eq:S-I0} & \LeanDecl{besselMajorant};
\LeanDecl{modifiedBesselI0Series}
(\emph{FiniteBesselMajorant.lean}; \emph{BesselI0Series.lean}) \\
\eqref{eq:S-term-majorant} & \LeanDecl{supplementalTermMajorant}
(\emph{FiniteBesselMajorant.lean}) \\
\eqref{eq:S-B-I0} &
\LeanDecl{besselMajorant_eq_exp_mul_modifiedBesselI0Series}
(\emph{BesselI0Identity.lean}) \\
\eqref{eq:S-coefficient-identity} &
\LeanDecl{one_add_twoX_add_X_sq_pow}
(\emph{BesselI0Coefficient.lean}) \\
\eqref{eq:S-finite-chain} &
\LeanDecl{prlEquation_finiteBounds};
\LeanDecl{prlEquation_besselMajorant}
(\emph{PRLEquations.lean}) \\
\eqref{eq:S-ratio-chain} &
\LeanDecl{baseline_div_besselMajorant_le_gramSecondMomentRatio};
\LeanDecl{baseline_mul_exp_neg_le_gramSecondMomentRatio};
\LeanDecl{gramSecondMomentRatio_le_baseline}
(\emph{FiniteBesselMajorant.lean}; \emph{FiniteExponentialBound.lean};
\emph{FiniteSum.lean}) \\
\eqref{eq:S-one-term-lower} and \eqref{eq:S-geometric-certificate} &
\LeanDecl{finiteTerm_factorial_lower_certificate};
\LeanDecl{literalGaussianGramHafnian_ratio_le_two_pow_neg_of_geometric_certificate}
(\emph{LogBoundaryConverse.lean}; \emph{Final.lean}) \\
\eqref{eq:S-y-limit} and \eqref{eq:S-general-B-limit} &
\LeanDecl{tendsto_gramHafnianFactor_besselMajorant}
(\emph{PRLEquations.lean}) \\
\eqref{eq:S-recurrence} & \LeanDecl{finiteTerm_succ_ratio}
(\emph{FiniteSum.lean}) \\
\eqref{eq:S-Tannery-bound} &
\LeanDecl{extendedFiniteTerm_besselSeriesTerm_bounds}
(\emph{AsymptoticCritical.lean}) \\
\eqref{eq:S-quadratic-F} &
\LeanDecl{prlEquation_besselCrossover}
(\emph{PRLEquations.lean}) \\
\eqref{eq:S-central-Stirling} &
\LeanDecl{tendsto_prlCentralBaseline_mul_sqrt_pi}
(\emph{PRLEquations.lean}) \\
\eqref{eq:S-quadratic-R} &
\LeanDecl{literalGaussianGramHafnian_ratio_quadratic_crossover_of_aspect}
(\emph{PRLBoundary.lean}) \\
\eqref{eq:S-superquadratic} &
\LeanDecl{tendsto_gramHafnianFactor_one_of_quadratic_ratio_zero};
\LeanDecl{literalGaussianGramHafnian_ratio_superquadratic_baseline}
(\emph{PRLEquations.lean}; \emph{Final.lean}) \\
\eqref{eq:S-weak-def} & \LeanDecl{HasWeakAntiConcentration}
(\emph{AsymptoticWeakAC.lean}) \\
\eqref{eq:S-log-sufficient-pointwise} and \eqref{eq:S-log-sufficient} &
\LeanDecl{one_div_two_mul_pow_le_gramSecondMomentRatio_of_log_scale};
\LeanDecl{literalGaussianGramHafnian_hasWeakAntiConcentration_of_log_scale}
(\emph{AsymptoticLogBoundary.lean}; \emph{Final.lean}) \\
\eqref{eq:S-beta-positive} and \eqref{eq:S-beta-lower} &
\LeanDecl{eventually_explicit_lower_bound_of_log_dimension_ratio_tendsto_pos};
\LeanDecl{literalGaussianGramHafnian_hasWeakAntiConcentration_of_log_dimension_ratio_tendsto_pos}
(\emph{PRLBoundary.lean}) \\
\eqref{eq:S-log-witness} & \LeanDecl{logarithmicWitness};
\LeanDecl{pow_le_two_pow_logarithmicWitness}
(\emph{LogBoundaryConverse.lean}) \\
\eqref{eq:S-certificate-limit} &
\LeanDecl{tendsto_geometricCertificateRatio_logarithmicWitness}
(\emph{LogBoundaryConverse.lean}) \\
\eqref{eq:S-o-regime} and \eqref{eq:S-log-converse} &
\LeanDecl{literalGaussianGramHafnian_not_hasWeakAntiConcentration_of_log_dimension_ratio_zero}
(\emph{Final.lean}) \\
\eqref{eq:S-rank-one} & \LeanDecl{gramHafnianFactor_one};
\LeanDecl{gramSecondMomentRatio_one};
\LeanDecl{literalGaussianGramHafnian_not_hasWeakAntiConcentration_rank_one}
(\emph{PRLEquations.lean}; \emph{FiniteSum.lean}; \emph{Final.lean}) \\
\bottomrule
\end{longtable}
\endgroup

\section{Formal verification and reproducibility}

The archived Lean 4 development
\cite{deMouraUllrich2021,Mathlib2020,ZhaoLeanVerification2026} provides an
explicit declaration or theorem specialization for all fourteen numbered
equations in the Letter and all fifty numbered equations in this Supplemental
Material. Every mathematical result asserted in this work within the Gaussian
surrogate scope is proved by a named Lean theorem and checked by the Lean
kernel. Definitions and
notational equations are checked by elaboration and mapped to their defining
declarations. The release pins the Lean and mathlib versions and provides
source, build instructions, a theorem crosswalk, and separate axiom audit entry
points for the Letter and Supplemental Material. The audited endpoints report
only propositional extensionality, classical choice, and quotient soundness. A
source scan of the audited Gaussian Gram hafnian tree found no \texttt{sorry},
\texttt{admit}, or project specific \texttt{axiom}.
Equation~\eqref{eq:S-conditional} is an
almost sure identity for $k\ge2$, while the integrated $k=1$ moment is verified
separately. The verification excludes finite Haar transfer, full small ball
anticoncentration, cited literature, figure generation, and physical
interpretation; none of these is a mathematical result claimed by
the paper.

\end{document}